\documentclass[conference,letterpaper]{IEEEtran}

\usepackage[top=0.75in, bottom=1.05in, left=0.75in, right=0.75in, heightrounded]{geometry}
\usepackage{setspace}
\usepackage{amsmath} 

\usepackage[hyphens]{url}
\usepackage{hyperref}
\hypersetup{colorlinks,allcolors=blue}

\usepackage[backend=bibtex,style=ieee,citestyle=ieee-comp]{biblatex}
\usepackage{comment}
\IEEEoverridecommandlockouts

\usepackage{graphicx} 
\usepackage{xcolor} 
\usepackage{minted} 
\usepackage{doi}
\usepackage{orcidlink}
\def\BibTeX{{\rm B\kern-.05em{\sc i\kern-.025em b}\kern-.08em
    T\kern-.1667em\lower.7ex\hbox{E}\kern-.125emX}}

\begin{document}

\title{Gravity Falls: A Comparative Analysis of Domain-Generation Algorithm (DGA) Detection Methods for Mobile Device Spearphishing
\thanks{Disclaimer: The views expressed are those of the authors and do not necessarily reflect the official policy or position of the U.S. Department of Defense or the U.S. Government. References to external sites do not constitute endorsement. Cleared for release on 24 FEB 2026 (DOPSR 26-T-0771). Claude AI was used to expedite tool development and data formatting.}
}

\author{\IEEEauthorblockN{Adam Dorian Wong\,\orcidlink{0009-0000-1832-6859}, John D. Hastings\,\orcidlink{0000-0003-0871-3622}}
\IEEEauthorblockA{\textit{The Beacom College of Computer \& Cyber Sciences} \\
\textit{Dakota State University}\\
Madison, SD, USA \\
adam.wong@trojans.dsu.edu, john.hastings@dsu.edu
}
}

\maketitle

\begin{abstract} 
Mobile devices are frequent targets of eCrime threat actors through SMS spearphishing (smishing) links that leverage Domain Generation Algorithms (DGA) to rotate hostile infrastructure, avoid individual domain blocks, and bypass perimeter enterprise defenses. Despite this, DGA research and evaluation largely emphasize malware C2 and email phishing datasets, leaving limited evidence on how well detectors generalize to smishing-driven domain tactics outside enterprise perimeters. This work addresses that gap by evaluating traditional and machine-learning DGA detectors against Gravity Falls, a new dataset derived from smishing links delivered between 2022 and 2025. Gravity Falls captures a single threat actor's evolution across four technique clusters, shifting from short randomized strings to dictionary concatenation and themed combo-squatting variants used for credential theft and fee/fine fraud. Two string-analysis approaches (Shannon entropy and Exp0se) and two ML-based detectors (an LSTM classifier and COSSAS DGAD) are assessed using Top-1M domains as benign baselines. Results are strongly tactic-dependent: performance is highest on randomized-string domains but drops on dictionary concatenation and themed combo-squatting, with generally low recall across multiple tool/cluster pairings. Overall, both traditional heuristics and some common ML detection methods are ill-suited for consistently evolving DGA tactics observed in Gravity Falls, motivating more context-aware approaches and providing a reproducible benchmark for future evaluation.
\end{abstract}


\section{Introduction}
The Domain Name System (DNS) protocol is often called the backbone of the Internet. Adversaries leverage Domain Generation Algorithms (DGA) to register fully-qualified domain names (FQDNs) at scale, creating mass ephemeral infrastructure for the purposes of command-and-control (C2), data theft, or data exfiltration. Threat actors rely upon DGA to rotate infrastructure when discovered. This perpetuates the scam campaigns through targeting, data collection, data theft, or monetary theft through resilient orchestration and bulletproof hosting.

Security discussions often treat users as the weakest link, but that assumption implicitly places the user inside an organization's defended perimeter. In SMS spearphishing (smishing), the target is frequently a private individual operating outside enterprise security boundaries, with limited controls and fewer warning signals. In this setting, adversaries can bypass institutional defenses while rotating domains rapidly and with ease; blocking DGA-driven infrastructure can therefore inhibit key stages of the adversary kill chain by degrading beaconing and operational continuity. Personal devices of users are not always authorized on the corporate network or protected by it. Threat actors are aware of this issue and exploit societal soft-targets for financial advantages.

Through text messages, eCrime groups can reach individuals directly, circumvent enterprise defenses like firewalls and forward web proxies. These conditions make users soft-targets on personal mobile devices. To mitigate SMS attacks, users have limited recourse beyond blocking phone numbers, reporting a text message, or deleting a text message. DGA provides adversaries with the ability to sustain SMS targeting operations and avoid blocklists. Varying DGA patterns allow adversaries to complicate detection efforts and reduce likelihood that individual domains are burned in discovery by the cybersecurity community.

In an effort to match the scale of attacks and provide some security for the public, the Gravity Falls project was founded to provide threat intelligence indicators and to disrupt hostile infrastructure. The underlying assumption is congruent with David Bianco's \textit{Pyramid-of-Pain} model of incident response - that suggests different indicators-of-compromise (IOCs) hold different weights to defenders and attackers. The initial intent was to allow users to incorporate these indicators for protecting their personal devices on their home network through simple DNS sinkhole solutions like Pi-Hole.

Motivated by smishing-driven infrastructure, Gravity Falls represents the collection of spearphishing text messages with consistent thematic changes to tactics, techniques, \& procedures (TTPs) and perceived attack motivations. Additionally, it represents a significant concerted effort and resource investment to deploy hostile infrastructure across the Internet. Domains were derived from hyperlinked URLs delivered through SMS text messages, iMessage (Apple), or email-to-SMS between 2022 and 2025, and organized by perceived theme.

This quantitative research experiment aims to test the Gravity Falls dataset against several tools and compare results. The desired end-state is to advise on DGA detector tools that best challenge specific DGA TTPs. Specifically, Gravity Falls encompasses four distinct clusters that reflect an annual evolution in the threat actor's technique from randomized strings (2022), to dictionary concatenation (2023), and then to themed combo-squatting variants (2024-2025).

This paper makes two primary contributions:
\begin{enumerate}
    \item Provides a new DGA dataset which is not yet widely known or incorporated into DGA detectors.
    \item Finds that traditional and ML methods are ill-suited to identifying the majority of malicious domains from the Gravity Falls dataset.
\end{enumerate}

The remainder of this paper frames the smishing literature review \& research gap, introduces the Gravity Falls dataset and experimental methodology, reports comparative detector results, and discusses implications, future work, and limitations.

\section{Related Work}
Prior efforts analyzing DGA activity often arose from high-profile malware operations such as SUNBURST \cite{noauthor_mar-10318845-1v1_2021}. A key distinction in that work was applying Shannon entropy to the subdomain/hostname (prefix) portion of an FQDN \cite{wong_detecting_2023, wong_malwaremorghulissunburst_2024}, and comparing subdomain entropy values to detect DNS-based data movement behaviors observed in SUNBURST C2 activity \cite{noauthor_mar-10318845-1v1_2021,wong_detecting_2023, wong_malwaremorghulissunburst_2024}. In contrast, Gravity Falls represents a multi-year SMS spearphishing dataset collected over four years for sinkholing and hobbyist threat hunting; an external report is cited as the key enabling attribution of the SMS campaign since 2022 \cite{krebs_china-based_2025}.

More broadly, surveys and meta-analyses have emphasized both the breadth of DNS misuse and the persistent gaps in detection. Zhauniarovich \cite{zhauniarovich_survey_2018} surveyed DNS abuse, collection methods, and analytic approaches for DNS-based attacks. Gardiner \cite{gardiner_security_2016} surveyed C2 tracking approaches spanning signatures, DNS transactions, and ML clustering, and highlighted that modern ML practices can be undermined by attacks against ML models, cautioning against blind-trust in AI/ML. Metcalf \cite{metcalf_ecosystem_2021} evaluated multiple DGA detection algorithms and concluded that a generalized detector is impossible.

Beyond DNS-only approaches, botnet countermeasures and industry methods extend to sinkholing and server seizure, while also noting the lack of economic controls \cite{georgoulias_botnet_2023}, and alternative detection signals outside of DNS such as transactional hashes in TLS handshakes \cite{noauthor_jarm_nodate, noauthor_salesforceja3_2025}.

Several lines of work focus specifically on string- and content-based domain analysis. Satoh \cite{satoh_superficial_2020} proposed ``superficial analysis'' and training-free approaches using web searches, whitelists, long-subdomain filtering, dictionary words, and web search reduction, but noted that adversaries can shift to second-level domains or short subdomains to avoid long-subdomain heuristics.  Lee \cite{lee_detecting_2024} examined C2 domains via word substrings (similar to N-grams) in romanized Chinese domains [30], while Liang \cite{liang_hagdetector_2022} compared detectors across short- and long-string DGA domains using forward and reverse N-grams to account for length and pattern effects.  Dictionary-focused work includes Liew's \cite{liew_use_2023} analysis of dictionary concatenations with selective routing to different ML models.

Finally, DGA detection has increasingly emphasized AI/ML methods. Chen \cite{chen_dga-based_2021} used LSTM feature extraction for DGA while leveraging Particle Swarm Optimization (PSO) to dynamically configure sampling. Suryotrisongko \cite{suryotrisongko_robust_2022} supplemented statistical ML with OSINT to establish confidence in maliciousness. Yang \cite{yang_detecting_2020} utilized heterogeneous deep neural networks (HDNN) and LSTM to assess strings forward and backward.  While the trend leans toward AI/ML solutions, organizations are often adopting AI/ML through prompt interfaces (e.g., Claude, ChatGPT, Perplexity) \cite{eckels_sunburst_nodate, guptta_modeling_2024, hassaoui_unsupervised_2024, thomson_novel_2025}, leaving an observed gap in actionable detection direction despite the diversity of DGA techniques.

\section{Research Gap} During the literature review process, it was found that a major focus was on malware or email phishing. However, there is not as much emphasis on SMS-based phishing. 
\begin{enumerate}
    \item First, one gap is that previous DGA research predominantly focuses on known malware C2 domains or datasets, and re-analyzing already widely-known data.
    \item Second, the Gravity Falls dataset represents several clusters of varying TTPs and consistent evolution of the DGA, by a real threat actor.
    \item Third, network security research heavily focuses on enterprise or industrial networks, but not necessarily defense of vulnerable mobile telecommunications infrastructure and individuals susceptible to eCrime.
    \item Fourth, there is literature in performance of a single detector and a dataset (one-to-one, or one-to-many), but not many publications on comparing multiple DGA techniques against multiple tools (many-to-many).
\end{enumerate}

The overall research objective is to determine how effective traditional and modern DGA detection techniques perform against relatively newer DGA data and the TTPs under the dataset dubbed: Gravity Falls.

\section{Dataset and Experimental Setup}
Gravity Falls \cite{wong2025gravity} is a dataset consisting of C2 domains delivered via SMS text messages between 2022 and 2025\footnote{The Gravity Falls dataset is publicly available (\doi{10.5281/zenodo.22003712}). Supporting code and supplemental material are available at: \url{https://github.com/MalwareMorghulis/GravityFalls}}.

The dataset is organized into four technique clusters that generally emerged within a calendar year; overall, the dataset is composed of two branches. The explicit blocklists contain observed and registered hostile domains and the second blocklist contains a list of predicted regex patterns. Together, these two branches are used for sinkholing. For the purposes of experimentation, only the observed domains were used in testing.

Between 2022-2025, each calendar year yielded an evolution or variation of the SMS spearphishing attack which suggests the threat actor modified their targeting capabilities, thematic content, and domain generation algorithm. These clusters were likely from the same threat actor due to unique behavioral patterns in delivery and commonalities in patterns-of-life such as: spoofed phone numbers or email-to-SMS senders, lure reuse, Uniform Resource Identifier (URI) pattern reuse, consistent writing styles, delivering within 24 hours of registration, infrastructure reuse on platforms like NameCheap, Cloudflare (AS13335), Alibaba servers, unusual Top-Level Domains (TLDs) or generic TLDs (gTLDs), combo-squatting patterns, and perceived focus on US citizens.

Over time, collection goals shifted from immediate response (e.g., Pi-hole blocklists and sinkholing) toward research-grade characterization as volume and variability increased; later waves required partial regex-based characterization for sinkholing when exact enumeration became impractical. In addition to direct observation, open-source reporting and infrastructure analysis (including Iris Investigate link-graph pivots) support the working attribution that this activity is consistent with the ``Smishing Triad'' ecosystem \cite{krebs_china-based_2025} described in public reporting.

The four experimental clusters evaluated in this paper are:

\begin{enumerate}
    \renewcommand{\arraystretch}{0.90}
    \item Cats Cradle (2022): 
    \begin{itemize}
        \item Perceived Technique: randomized letters, use of randomized alphabetical characters within 5-8 characters. 
        \item Assessed Purpose: target validation through fake CAPTCHA.
    \end{itemize}
    \item Double Helix (2023):
        \begin{itemize}
            \item Perceived Technique: dual words, use of dictionary wordlist concatenation.
            \item Assessed Purpose: target validation through fake CAPTCHA.
        \end{itemize}
    \item Pandora's Box (2024): 
        \begin{itemize}
            \item Perceived Technique: postal-theme (package delivery, Amazon, USPS, FedEx, etc) spearphishing URLs. 
            \item Assessed Purpose: credential and identity theft.
        \end{itemize}
    \item Easy Rider (2025): 
        \begin{itemize}
            \item  Perceived Technique: toll and government-theme (DMV, Speeding Fines, EzPass, etc.) phishing URLs.
            \item Assessed Purpose: commit fraud through fake fees or fines.
        \end{itemize}
\end{enumerate}

This research threat model assumed that the group behind these DGA indicators was well-resourced and technically-proficient. This group was capable of registering thousands of domains and sending SMS texts from email addresses (legitimate or spoofed) and phone numbers (domestic or international) and heavily favor registrars with weaker restrictions as well as Chinese cloud infrastructure. This group knew enough English to craft messages of urgency, but not quite enough for regional nuances. The threat actors did not appear to be native or local citizens because text message content varied and sometimes used incorrect government titles (e.g, ``Bureau'' instead of ``Department''). The threat actors did not appear to focus on intelligence collection of targets or relied on disruptive actions against infrastructure or personal devices. This group targeted civilians away from enterprise-grade protections such as firewalls and intrusion detection systems. The initial entry vector for attacks included spearphishing text messages with a hostile domain and accompanying specific Uniform Resource Identfier (URI). The motive and objectives of this group appeared to have evolved over time from target validation to financial scams.

As of August 2026, the majority of the original dataset was recovered and standardized through retroactive searches in passive DNS (pDNS) through a single commercial vendor: DomainTools. These pDNS record searches were time-boxed around the approximate dates in which first-party direct hostile messages were sent to members of the research team. Additional SMS samples were crowd-sourced through voluntary submissions from fellow information security practitioners who were aware of Gravity Falls (and who were also targeted victims). These crowd-sourced samples were inspected for text message content relevance, URL pattern, and domain registration similarities before including in the threat intelligence pivots while searching for adjacent infrastructure.

The control groups were 3,000 samples each randomly selected from four major Top-1M lists: Alexa Top-1M (2017), a static list for common US web services; Cisco Top-1M (2025), a dynamic list for various Internet services; Cloudflare Top-1M (2025), a dynamic list for global web services; and Majestic Top-1M (2025), a dynamic list for UK web services.

Each experimental group consisted of 3,000 domains from the respective activity cluster. These hostile domains were sampled randomly without replacement. The full Gravity Falls git repo was de-duplicated successfully, so domains reappearing in the spam campaign were removed and predicted regex patterns separated from real observed domains.

For the overall production blocklist, subdomains were treated as separate entities (e.g., evil.com vs. something.evil.com). During this research effort, they were treated as identical and removed during sample screening. So, two sets of de-duplication occured, at the repo-level and at the sampling group-level.

Following precedent in prior DGA literature \cite{vinayakumar_evaluating_2018, highnam_real-time_2021, tuan_utl_dga22_2023, maia_end--end_2024, noauthor_endgameincdga_predict_2025}, Top-1M domains were treated as generally benign for the purpose of measuring deviations under a consistent baseline assumption; this risk is accepted because manual non-random sampling for a control group would be inappropriate. Alexa was also used as a baseline because it is generally static, while Cisco reflects wider internet infrastructure and FQDNs, Majestic provides a UK perspective, and Cloudflare provides a global perspective. 

\subsection{Cluster Characteristics}
The following subsections summarize the lexical and operational characteristics observed within each cluster.

\subsubsection{Cats Cradle (2022)}
Early activity primarily used short randomized strings (often ~7 characters) paired with common or generic TLDs (e.g., *.com) and occasional newer gTLDs (e.g., *.online). Character distribution appeared relatively uniform, consistent with deliberate randomness, and collection in this period relied on exact observed domains rather than pattern-based prediction. In multiple cases, the landing pages resembled fake CAPTCHA portals, consistent with a ``target validation'' step. 

\subsubsection{Double Helix (2023)}
The following year, the threat actor applied dictionary/wordlist concatenation in both second-level labels and, in some cases, word-like TLD choices. The increased ``word-looking'' structure makes these domains harder to spot by randomness-oriented detectors without additional context. The corpus also showed broader use of newer IANA gTLDs (e.g., *.business, *.finance). Occasional truncations were observed; a plausible hypothesis is a memory/encoding constraint in the generation pipeline (not proven, but consistent with the artifacts). Fake CAPTCHA-style pages persisted as a likely validation mechanism. 

\subsubsection{Pandora's Box (2024)}
Messaging content matured into more professional package-delivery lures. During this period, Apple modified iMessage (disabled hyperlinks from unknown senders unless the recipient responds \cite{martin_how_2025}). This likely shifted attacker behavior toward interaction-driven activation. Toward late 2024, the campaign  used ``@'' username redirection in URLs with combo-squatting \cite{choudhary_cybercriminals_2024}. Infrastructure and naming patterns increased in sophistication: keyword splitting between subdomain and domain (combo-squatting), plus postpended random suffixes that complicate sinkholing by obscuring stable patterns (e.g., ``track'' + random tail). This cluster heavily favored Chinese-based TLDs/infrastructure and blended dictionary terms, brand tokens, and small randomization. 

\subsubsection{Easy Rider (2025)}
By early 2025, the DGA technique stabilized into themed combo-squatting with brand/keyword associations plus randomized suffixes, but morphed lures around government/toll themes (e.g., DMV, state police, parking/tolls, EZ-Pass), leveraging fear of penalties rather than delivery delays. Nuanced spelling mistakes for title names like ``Department of...'' vs. ``Bureau of...'' further supported that the attackers are likely not familiar with US State, Local, Tribal, \& Territorial (SLTT) vernaculars and are likely not US citizens. Delivery infrastructure also evolved: email-to-iMessage/email-to-SMS became prevalent and the campaign began using foreign numbers outside the US (+1), increasing sender ephemerality.

Although beyond the scope of research, Gravity Falls represents a continuous change of TTPs over the span of several years. Each successive year demonstrated a variation in perceived desired end states and updates to domain generation capabilities. Infrastructure use varied across different VPS hosting providers and registrars demonstrating the sheer scale of operations.

\subsection{Collection Workflow}
Collection began from recipient-side observation of smishing messages and extraction of hyperlinked URLs/domains. 

\textbf{2022-2024:} The hostile domain would be submitted to a WHOIS service provider such as DomainTools. The IP address would be recorded although other metadata such as registrant, geolocation, and URI were inconsistently recorded. This IP address would be searched in passive DNS records through vendors like RecordedFuture's SecurityTrails. For context, URLscan was also used to see snapshots of C2 domains for visual context. SecurityTrails provided free but heavily limited access which adjacent domains were identified. This process was heavily dependent on several tools.

\textbf{2024-2025:} In collaboration with DomainTools, the data collection method changed towards use of their commercial platforms called: ``IRIS Investigate'' and ``DNSDB Scout''. This commercial access provided additional and enhanced capabilities such as node-link graph visualizations, historical WHOIS records, passive DNS, and data extraction to CSV or JSON files.

Separately, indicators were shared via LevelBlue/AlienVault OTX before indicator migration to GitHub due to OTX search limitations; during migration, practical issues were observed around public IOC posting (e.g., automated account suspensions when email IOCs appear in repos), reinforcing need to curate what is shared publicly.

\textbf{2026:} With the assistance of DomainTools, pDNS records were recovered and standardized for the early years spanning 2022-2024. Indicators in the Gravity Falls repo were finally normalized. Datasets were collected consistently, uniformly, and at scale from a reliable cyber threat intelligence research partner. Effectively, data collection started and ended in one set of vendor tools.

\section{Methodology}
\subsection{Detector Toolsets}
This study evaluated two traditional string-analysis approaches and two ML-based detectors.

\textit{Shannon entropy} \cite{shannon_mathematical_1948} quantifies information content within a domain string. 
Alexa Top-1M was chosen because this list was static, but still relevant. It focused on consumer-facing web services rather than a multitude of services (cloud or router FQDNs). These properties made it suitable for use in character counting.
In this experiment, entropy values were calculated using a character probability table derived from Alexa Top-100K and sampled for consistency across groups. 
While counting characters, subdomains (if any) and TLDs (including country codes) were ignored because the experiment focused on the domain name or rather the portion before the TLD. Since these Top-1M lists use scaled ordinal ranking, these Top-1M lists do not always explicitly outline the number of site-visitations per domain. Therefore, the character probability table was calculated based on the first 100K domains or Alexa Top-100K domains to take ordinal scaling into consideration. Ignoring subdomains (which were uncommon in the Alexa list) and TLDs was a design decision to avoid skewing the probability table based on prefixes and TLDs (e.g., www.* or *.com). 

Entropy is defined as:
\begin{equation}
    H(x) = -\sum_{i=1}^{n} p(x_i) \log_2(p(x_i))
\end{equation}

\textit{Exp0se DGA Detector} \cite{starr_exp0sedga_detector_2025} is a traditional detector based on domain string characteristics (entropy, consonant count, and string length thresholds). One weakness is that this tool overlooks short domains.

\textit{MiaWallace0618 DGA Detection} \cite{qu_miawallace0618dga_detection_2024} applies an LSTM using one-hot encoding of TLDs to classify domains.

\textit{COSSAS DGAD} (Community of Open-Source Security Automation Software DGA Detective) \cite{noauthor_cossasdgad_2025} uses a Temporal Convolutional Network (TCN) trained on Shadowserver Foundation data.  DGAD produces both a substring word assessment and an overall domain assessment; this study reports overall domain performance for consistency across tools. DGAD was unable to assess domain strings longer than 120 characters in total length (including delimiters) which are common in Internet Service Provider (ISP) router hostnames.

All tools were executed on a Ubuntu 24.04.03 LTS Virtual Machine (VM); containerized tools were run via Docker. Experiment replication steps can be found on the Gravity Falls Github repo.

\renewcommand{\arraystretch}{0.90}
\begin{table*}[ht]
\centering
\caption{Performance summary by experimental group (Accuracy)}

\label{tab:perf_summary}
\begin{tabular}{lcccc}
\hline
\textbf{Method} & \textbf{Cats Cradle} & \textbf{Double Helix} & \textbf{Pandora's Box} & \textbf{Easy Rider} \\
\hline
Shannon Entropy &
0.0503 &
0.0070 &
0.0687 &
0.0017 \\
Exp0se (static signatures) &
0.9173 &
0.0147 &
0.3983 &
0.4673 \\
LSTM (miaWallace0618) &
0.0017 &
0.0040 &
0.0637 &
0.1083 \\
DGAD (overall domain) &
0.9117 &
0.0443 &
0.0017 &
0.0060 \\
\hline
\end{tabular}
\end{table*}

\subsection{Comparative Evaluation \& Metrics}
Tool outputs were curated into spreadsheets for comparative analysis with an expected binary classification. True Positive (TP) represents correctly identified hostile domains, and True Negative (TN) represents legitimate domains identified as benign. False Positive (FP) represents a benign domain misidentified as DGA as benign, and False Negative (FN) represents a hostile domain misidentified as benign. The confusion matrix is based on the individual tool's ability to detect the hostile domain. If a tool was unable to process a domain, then the domain was considered a FN result akin to a security solution in fail-open state.
\[
\small
\begin{array}{c|cccc}
\text{Verdict} & \text{Shannon} & \text{Exp0se} & \text{Mia} & \text{DGAD} \\
\hline
\text{TP} & \text{suspicious} & \text{is\_DGA:TRUE} & \text{dga} & \text{is\_DGA:TRUE} \\
\text{TN} & \text{normal} & \text{is\_DGA:FALSE} & \text{legit} & \text{is\_DGA:FALSE} \\
\text{FP} & \text{suspicious} & \text{is\_DGA:TRUE} & \text{dga} & \text{is\_DGA:TRUE} \\
\text{FN} & \text{normal} & \text{is\_DGA:FALSE} & \text{legit} & \text{is\_DGA:FALSE} \\
\end{array}
\]\\

Precision, accuracy, and recall using standard definitions are below:
\begin{equation}
    Precision = TP/(TP+FP)
\end{equation}
\begin{equation}
    Accuracy = (TP+TN)/(TP+TN+FP+FN)
\end{equation}
\begin{equation}
    Recall = TP/(TP+FN)
\end{equation}

Accuracy was the predominant metric for experimentation. Due to the binary categorization and limited usefulness, precision was omitted. Recall ratios matched corresponding accuracies.

\section{Results}
This section summarizes detector performance on the four Gravity Falls experimental groups (Cats Cradle, Double Helix, Pandora's Box, Easy Rider). Unless otherwise noted, the control domains are treated as benign Top-1M samples (see Methodology). The full results can be viewed in the Gravity Falls public repo.

Table \ref{tab:perf_summary} consolidates precision, accuracy, and recall across the evaluated methods. Overall, performance varied substantially by technique cluster, with the strongest results occurring on randomized-string domains (Cats Cradle) and notably weaker results on dictionary concatenation (Double Helix) and themed combo-squatting clusters (Pandora's Box, Easy Rider).

Across methods, Cats Cradle produced the clearest separation signal (highest precision and accuracy achieved by Exp0se and DGAD), while Double Helix remained difficult for every method tested. Detailed per-domain outputs and extended characterization tables are omitted here for space and are retained in the companion materials/repository.

\section{Discussion}
\subsection{Interpretation of Results}
Table \ref{tab:perf_summary} shows that detector efficacy depends strongly on the domain-generation tactic, not simply whether a domain is algorithmic. In particular, the evaluated methods achieved their strongest performance on randomized-string domains (Cats Cradle), while dictionary concatenation (Double Helix) and themed combo-squatting variants (Pandora's Box and Easy Rider) consistently degraded detection performance across both traditional and ML-based approaches.

Among traditional approaches, Exp0se was most successful on completely random strings (Cats Cradle) and only partially effective on the combo-squatting clusters (Pandora's Box, Easy Rider), while remaining weak on dictionary wordlists (Double Helix). This aligns with the design of string-heuristic detectors as high-throughput sieves for high-entropy or structurally suspicious domains, but suggests that campaigns combining recognizable tokens with short random postfixes can bypass such heuristics at scale. However, Exp0se relied upon arbitrary static characterization.

For ML-based detectors, both the LSTM tool and DGAD exhibited limited generalization to Gravity Falls techniques beyond Cats Cradle, despite being functionally consistent across reruns (i.e., no evidence of tool failure). In practice, this implies that defenders should treat strong DGA results reported on widely used malware-family corpora as insufficient evidence of robustness against smishing-driven domain tactics that blend dictionary words, brand tokens, and minor randomization. Although anecdotal, LLM-assisted inspection was able to identify a theme across clusters, suggesting a potential role for LLM augmentation in future detectors.

\subsection{Defensive Implications}
Gravity Falls demonstrates that a real-world SMS phishing campaign can deploy multiple domain-generation strategies over time and still evade both traditional and modern detectors under this evaluation.  For short-term defensive value, these results support a layered approach: use fast lexical heuristics to triage obvious randomized domains (Cats Cradle-like), but rely on additional context (message content, hosting/infrastructure signals, and brand/keyword abuse policies) when confronting dictionary and combo-squatting tactics (Double Helix/Pandora's/Easy Rider).

This paper focuses on the core methodology and the comparative performance results of the toolsets. Additional per-cluster characterization (e.g., extended TLD and string-length distributions) and qualitative artifacts (e.g., representative landing pages/messages) are intentionally deferred. These materials will be developed and detailed in a planned journal extension to provide deeper interpretability and richer examples, but are not required to support the quantitative findings reported here.

\subsection{Practical Takeaways for Defenders \& Replication}
Some lessons were learned for future consideration or improvements:
\begin{itemize}
    \item \textit{Claude AI}. Claude was fairly decent at detecting a common theme across the four clusters under Gravity Falls. Perhaps future DGA tools will incorporate Large Language Models (LLMs) to assist in scrutinizing domains for general trend analysis.
    \item \textit{Testing with DGA Datasets}. As observed in the Gravity Falls repo, domains can be generated a variety of ways. Detector tool testing needs to account for various generation methods.
    \item \textit{Deduplication of Samples}. Each tool approaches detection differently (subdomains vs. first-level domain). Deduplication checks should happen at each step of processing to avoid mistakes that delay testing. 
    \item \textit{Test Predicted Domains Separately}. Prior iterations of this experiment used a mixed experimental set of both real and predicted hostile domains. Current iterations of testing segmented real and predicted regexes while only testing hostile domains, not predicted ones. Sample sizes should also be increased in future iterations.
\end{itemize}

\section{Future Work}
With access to DomainTools, WHOIS and domain data collection is retroactively searched, standardized, and exported with exact matches in tabular form, for higher fidelity data. The experiment could also be duplicated with a carefully curated benign list from Alexa, Cisco, Cloudflare, or Majestic rather than as-is. Additionally, newer ML techniques will be tested against Gravity Falls to improve understanding of current findings. Further research will be conducted around the threat actors conducting these mobile spearphishing and DGA registration operations.

\section{Limitations}
This study has several limitations that affect internal and external validity. First, Gravity Falls indicators were not collected consistently at the start. Efforts have been made in 2026 to normalize the data. It is possible that observed data in SecurityTrails (in 2022) may not reflect what DomainTools has currently (2026) and vice versa.

Second, sampling size was reduced from 10K of mixed predicted-observed domains (version 1 testing) to 3K distinctly hostile domains (version 2 testing). This change occurred in recent iterations based on available time, renewed data curation, and prior peer-reviewer recommendations to adjust the experimental design.

Third, the control and experimental groups are seen as binary in nature. Top-1M lists are just as likely to have real hostile domains. In this effort, Control and Experimental Groups were viewed as all-benign or all-hostile.

Fourth, the benign baseline is imperfect. Alexa Top-1M is static and outdated and should not be considered fully benign, even if prior research has used it as a benign proxy; it was selected here because it is freely available and stable. 

Fifth, several evaluated tools are older implementations. The detector tools are not bleeding-edge and some public tools lacked updates or supporting documentation. These tools were selected because they were directly available for experimentation and for comparing traditional and ML methods; therefore, results reflect the specific tool versions tested.

Finally, precision and recall ratios are limited in validity and usefulness due to the binary categorization of domains. Although informative, accuracy had limited operational transferability, especially considering the variety of detectors tested.
\\
\section{Conclusion}
This work introduced Gravity Falls, a new dataset of SMS spearphishing-delivered domains collected between 2022 and 2025 and organized into four DGA technique clusters reflecting an evolution from randomized strings to dictionary concatenation and themed combo-squatting variants. Across a many-to-many evaluation of traditional and ML-based detectors, results show that detector performance is highly tactic-dependent: methods performed best on randomized domains (Cats Cradle) and degraded substantially on dictionary concatenation (Double Helix) and themed combo-squatting clusters (Pandora's Box, Easy Rider). These findings suggest that reported strong DGA performance on commonly used corpora may not generalize to smishing-driven tactics that blend dictionary tokens, brand terms, and minor randomization. Limitations include the dataset's efforts to standardize collection, reduced sample size, use of Top-1M as benign baselines, and need for testing against more modern ML algorithms, which should be addressed in future extensions.


\printbibliography

@software{qu_miawallace0618dga_detection_2024,
	title = {{miaWallace}0618/{DGA}\_Detection},
	url = {https://github.com/miaWallace0618/DGA_Detection},
	author = {Qu, Rachel},
	urldate = {2025-11-19},
	date = {2024-12-28},
	note = {original-date: 2019-01-29T15:55:03Z},
}

@article{shannon_mathematical_1948,
	title = {A {Mathematical} {Theory} of {Communication}},
	volume = {27},
	url = {https://people.math.harvard.edu/~ctm/home/text/others/shannon/entropy/entropy.pdf},
	language = {en},
	journal = {The Bell System Technical Journal},
	author = {Shannon, C E},
	month = oct,
	year = {1948},
	pages = {379--423},
}

@article{highnam_real-time_2021,
	title = {Real-Time Detection of Dictionary {DGA} Network Traffic Using Deep Learning},
	volume = {2},
	@issn = {2661-8907},
	@url = {https://doi.org/10.1007/s42979-021-00507-w},
	doi = {10.1007/s42979-021-00507-w},
	abstract = {Botnets and malware continue to avoid detection by static rule engines when using domain generation algorithms ({DGAs}) for callouts to unique, dynamically generated web addresses. Common {DGA} detection techniques fail to reliably detect {DGA} variants that combine random dictionary words to create domain names that closely mirror legitimate domains. To combat this, we created a novel hybrid neural network, Bilbo the “bagging” model, that analyses domains and scores the likelihood they are generated by such algorithms and therefore are potentially malicious. Bilbo is the first parallel usage of a convolutional neural network ({CNN}) and a long short-term memory ({LSTM}) network for {DGA} detection. Our unique architecture is found to be the most consistent in performance in terms of {AUC}, \$\$F\_1\$\$score, and accuracy when generalising across different dictionary {DGA} classification tasks compared to current state-of-the-art deep learning architectures. We validate using reverse-engineered dictionary {DGA} domains and detail our real-time implementation strategy for scoring real-world network logs within a large enterprise. In 4 h of actual network traffic, the model discovered at least five potential command-and-control networks that commercial vendor tools did not flag.},
	pages = {110},
	number = {2},
	journaltitle = {{SN} Computer Science},
	shortjournal = {{SN} {COMPUT}. {SCI}.},
	author = {Highnam, Kate and Puzio, Domenic and Luo, Song and Jennings, Nicholas R.},
	@urldate = {2025-11-16},
	@date = {2021-02-22},
    year={2021},
	langid = {english},
	keywords = {Botnets, Deep learning, Domain generation algorithm, Malware, Network security, Neural networks},
}

@article{maia_end--end_2024,
	title = {An end-to-end framework for private {DGA} detection as a service},
	volume = {19},
	@issn = {1932-6203},
	@url = {https://journals.plos.org/plosone/article?id=10.1371/journal.pone.0304476},
	doi = {10.1371/journal.pone.0304476},
	abstract = {Domain Generation Algorithms ({DGAs}) are used by malware to generate pseudorandom domain names to establish communication between infected bots and command and control servers. While {DGAs} can be detected by machine learning ({ML}) models with great accuracy, offering {DGA} detection as a service raises privacy concerns when requiring network administrators to disclose their {DNS} traffic to the service provider. The main scientific contribution of this paper is to propose the first end-to-end framework for privacy-preserving classification as a service of domain names into {DGA} (malicious) or non-{DGA} (benign) domains. Our framework achieves these goals by carefully designed protocols that combine two privacy-enhancing technologies ({PETs}), namely secure multi-party computation ({MPC}) and differential privacy ({DP}). Through {MPC}, our framework enables an enterprise network administrator to outsource the problem of classifying a {DNS} (Domain Name System) domain as {DGA} or non-{DGA} to an external organization without revealing any information about the domain name. Moreover, the service provider’s {ML} model used for {DGA} detection is never revealed to the network administrator. Furthermore, by using {DP}, we also ensure that the classification result cannot be used to learn information about individual entries of the training data. Finally, we leverage post-training float16 quantization of deep learning models in {MPC} to achieve efficient, secure {DGA} detection. We demonstrate that by using quantization achieves a significant speed-up, resulting in a 23\% to 42\% reduction in inference runtime without reducing accuracy using a three party secure computation protocol tolerating one corruption. Previous solutions are not end-to-end private, do not provide differential privacy guarantees for the model’s outputs, and assume that model embeddings are publicly known. Our best protocol in terms of accuracy runs in about 0.22s.},
	pages = {e0304476},
	number = {8},
	journaltitle = {{PLOS} {ONE}},
	shortjournal = {{PLOS} {ONE}},
	author = {Maia, Ricardo J. M. and Ray, Dustin and Pentyala, Sikha and Dowsley, Rafael and Cock, Martine De and Nascimento, Anderson C. A. and Jacobi, Ricardo},
	@urldate = {2025-11-16},
    year={2025},
	@date = {2024-08-28},
	langid = {english},
	publisher = {Public Library of Science},
	keywords = {Computer security, Convolution, Deep learning, Machine learning, Multilayer perceptrons, Neural networks, Neurons, Recurrent neural networks},
}

@article{vinayakumar_evaluating_2018,
	title = {Evaluating deep learning approaches to characterize and classify the {DGAs} at scale},
	volume = {34},
	@issn = {1064-1246},
	@url = {https://doi.org/10.3233/JIFS-169423},
	doi = {10.3233/JIFS-169423},
	abstract = {In recent years, domain generation algorithms ({DGAs}) are the foundational mechanisms for many malware families. Mainly, due to the fact that {DGA} can generate immense number of pseudo random domain names to associate to a command and control (C2) infrastructures. This paper focuses on to detect and classify the pseudo random domain names without relying on the feature engineering or any other linguistic, contextual or semantics and statistical information by adopting deep learning approaches. A deep learning approach is a complex model of traditional machine learning mechanism that has received renewed interest by solving the long-standing tasks in artificial intelligence ({AI}) related to the field of natural language processing, image recognition, speech processing and many others. They have immense capability to extract optimal feature representations by taking input as in the form of raw input texts. To leverage this and to transfer the performance enhancement in aforementioned areas towards characterize, detect and classify the {DGA} generated domain names to a specific malware family, this paper adopts deep learning mechanisms with a known one million benign domain names from Alexa, {OpenDNS} and a corpus of malicious domain names generated from 17 {DGA} malware families in real time for training in character and bigram level and a trained model has been evaluated on the {OSNIT} data set in real-time. Specifically, to understand the effectiveness of various deep learning mechanisms, we used recurrent neural network ({RNN}), identity-recurrent neural network (I-{RNN}), long short-term memory ({LSTM}), convolution neural network ({CNN}), and convolutional neural network-long short-term memory ({CNN}-{LSTM}) architectures. Additionally, to find out an optimal architecture, experiments are done with various configurations of network parameters and network structures. All experiments run up to 1000 epochs with a learning rate set in the range [0.01-0.5]. Overall, deep learning approaches, particularly family of recurrent neural network and a hybrid network (where the first layer is {CNN} and a subsequent layer is {LSTM}) have showed significant performance with a highest detection rate 0.9945 and 0.9879 respectively. The main reason is deep learning approaches have inherent mechanisms to capture hierarchical feature extraction and long range-dependencies in sequence inputs.},
	@pages = {1265--1276},
	number = {3},
	journaltitle = {Journal of Intelligent \& Fuzzy Systems},
	author = {Vinayakumar, R. and Soman, K.P. and Poornachandran, Prabaharan and Sachin Kumar, S.},
	@urldate = {2025-11-16},
	@date = {2018-03-22},
    year={2018},
	@note = {Publisher: {SAGE} Publications},
}

@article{tuan_utl_dga22_2023,
	title = {{UTL}\_DGA22 - a dataset for {DGA} botnet detection and classification},
	volume = {221},
	@issn = {1389-1286},
	@url = {https://www.sciencedirect.com/science/article/pii/S1389128622005424},
	doi = {10.1016/j.comnet.2022.109508},
	abstract = {The {DGA} botnet prevention is a burning topic in cybersecurity, with two problems: detection and classification. The {DGA} botnet dataset plays an essential role in the research allowing researchers to evaluate their proposed solutions. This study introduces a new dataset on {DGA} botnets named {UTL}\_DGA22. Our proposed dataset not only inherits previous datasets' results but also has got own advantages. First, our new dataset includes only domain records and no other raw network traffic, helping to address the {DGA} botnet problem. Second, we removed duplicated botnet {DGA} families and added new botnet families for a total of 76 {DGA} botnet families presented. Third, we propose a valuable set of attributes as input for classification algorithms. Our experiments using the proposed features with several machine learning algorithms have had good results. It shows that our proposed attributes are firmly suitable for the input of the {DGA} botnet solution. Finally, we carefully compiled the dataset and attribute description documents to make it easy for researchers to use. The {UTL}\_DGA22 dataset can serve as a database for researchers to develop their algorithms while objectively evaluating different solutions.},
	@pages = {109508},
	journaltitle = {Computer Networks},
	shortjournal = {Computer Networks},
	author = {Tuan, Tong Anh and Anh, Nguyen Viet and Luong, Tran Thi and Long, Hoang Viet},
	@urldate = {2025-11-16},
    year={2023},
	@date = {2023-02-01},
	keywords = {{DGA} botnet classification, {DGA} botnet detection, {DGA} botnets datasets, Feature extraction, Machine learning},
}

@online{martin_how_2025,
	title = {How to spot and block scam texts on your {iPhone}},
	url = {https://appleinsider.com/articles/25/01/12/psa-text-scammers-resorting-to-new-tactics-to-get-you-to-enable-phishing-links},
	titleaddon = {{AppleInsider}},
	author = {Martin, Charles},
	urldate = {2025-11-14},
	date = {2025-01-12},
	langid = {english},
}

@online{choudhary_cybercriminals_2024,
	title = {Cybercriminals are using "@" Symbol to Mask Phishing Links},
	url = {https://gurudattchoudhary.medium.com/how-cybercriminals-use-the-symbol-to-mask-phishing-links-7983d1028fd1},
	titleaddon = {Medium},
	author = {Choudhary, Gurudatt},
	urldate = {2025-11-14},
	date = {2024-08-18},
	langid = {english},
}

@software{starr_exp0sedga_detector_2025,
	title = {exp0se/dga\_detector},
	url = {https://github.com/exp0se/dga_detector},
	author = {Starr, {RIngo}},
	urldate = {2025-10-25},
	date = {2025-10-11},
	note = {original-date: 2015-11-15T10:44:11Z},
}

@software{noauthor_cossasdgad_2025,
	title = {{COSSAS}/dgad},
	rights = {Apache-2.0},
	url = {https://github.com/COSSAS/dgad},
	publisher = {{COSSAS}},
	urldate = {2025-09-20},
	date = {2025-08-13},
	note = {original-date: 2021-09-15T06:36:02Z},
}

@software{noauthor_endgameincdga_predict_2025,
	title = {endgameinc/dga\_predict},
	rights = {{GPL}-2.0},
	url = {https://github.com/endgameinc/dga_predict},
	publisher = {{ENDGAME}},
	urldate = {2025-09-19},
	date = {2025-09-03},
	note = {original-date: 2016-11-18T15:20:18Z},
}

@article{yang_detecting_2020,
	title = {Detecting Stealthy Domain Generation Algorithms Using Heterogeneous Deep Neural Network Framework},
	volume = {8},
	doi = {10.1109/ACCESS.2020.2988877},
	pages = {82876--82889},
	journaltitle = {{IEEE} Access},
	author = {Yang, Luhui and Liu, Guangjie and Dai, Yuewei and Wang, Jinwei and Zhai, Jiangtao},
	date = {2020},
}

@article{chen_dga-based_2021,
	title = {{DGA}-based botnet detection toward imbalanced multiclass learning},
	volume = {26},
	doi = {10.26599/TST.2020.9010021},
	pages = {387--402},
	number = {4},
	journaltitle = {Tsinghua Science and Technology},
	author = {Chen, Yijing and Pang, Bo and Shao, Guolin and Wen, Guozhu and Chen, Xingshu},
	date = {2021},
}

@article{satoh_superficial_2020,
	title = {A Superficial Analysis Approach for Identifying Malicious Domain Names Generated by {DGA} Malware},
	volume = {1},
	doi = {10.1109/OJCOMS.2020.3038704},
	pages = {1837--1849},
	journaltitle = {{IEEE} Open Journal of the Communications Society},
	author = {Satoh, Akihiro and Fukuda, Yutaka and Hayashi, Toyohiro and Kitagata, Gen},
	date = {2020},
}

@article{lee_detecting_2024,
	title = {Detecting Domain Names Generated by {DGAs} With Low False Positives in Chinese Domain Names},
	volume = {12},
	doi = {10.1109/ACCESS.2024.3454242},
	@pages = {123716--123730},
	journaltitle = {{IEEE} Access},
	author = {Lee, Huiju and Do Yoo, Jeong and Jeong, Seonghoon and Kim, Huy Kang},
	date = {2024},
	keywords = {Botnet, Chatbots, Deep learning, Detectors, Domain Name System, Feature extraction, Long short term memory, Malware, Servers, deep learning, domain generation algorithm, embedding, subword segmentation},
}

@article{suryotrisongko_robust_2022,
	title = {Robust Botnet {DGA} Detection: Blending {XAI} and {OSINT} for Cyber Threat Intelligence Sharing},
	volume = {10},
	doi = {10.1109/ACCESS.2022.3162588},
	pages = {34613--34624},
	journaltitle = {{IEEE} Access},
	author = {Suryotrisongko, Hatma and Musashi, Yasuo and Tsuneda, Akio and Sugitani, Kenichi},
	date = {2022},
}

@article{liang_hagdetector_2022,
	title = {{HAGDetector}: Heterogeneous {DGA} domain name detection model},
	volume = {120},
	@issn = {0167-4048},
	doi = {10.1016/j.cose.2022.102803},
	@journaltitle = {{COMPUTERS} \& {SECURITY}},
	journaltitle = {Computers \& Security},
	author = {Liang, Jianbing and Chen, Shuhui and Wei, Ziling and Zhao, Shuang and Zhao, Wei},
	@date = {2022-09},
    year={2022}
}

@article{liew_use_2023,
	title = {Use of subword tokenization for domain generation algorithm classification},
	volume = {6},
	@issn = {2523-3246},
	doi = {10.1186/s42400-023-00183-8},
	number = {1},
	journaltitle = {Cybersecurity},
	author = {Liew, Sea Ran Cleon and Law, Ngai Fong},
	@date = {2023-09-07},
    year={2023}
}

@article{gardiner_security_2016,
	title = {On the Security of Machine Learning in Malware C\&C Detection: A Survey},
	volume = {49},
	@issn = {0360-0300},
	@url = {https://doi.org/10.1145/3003816},
	doi = {10.1145/3003816},
	abstract = {One of the main challenges in security today is defending against malware attacks. As trends and anecdotal evidence show, preventing these attacks, regardless of their indiscriminate or targeted nature, has proven difficult: intrusions happen and devices get compromised, even at security-conscious organizations. As a consequence, an alternative line of work has focused on detecting and disrupting the individual steps that follow an initial compromise and are essential for the successful progression of the attack. In particular, several approaches and techniques have been proposed to identify the command and control (C8C) channel that a compromised system establishes to communicate with its controller.A major oversight of many of these detection techniques is the design’s resilience to evasion attempts by the well-motivated attacker. C8C detection techniques make widespread use of a machine learning ({ML}) component. Therefore, to analyze the evasion resilience of these detection techniques, we first systematize works in the field of C8C detection and then, using existing models from the literature, go on to systematize attacks against the {ML} components used in these approaches.},
	number = {3},
	journaltitle = {{ACM} Comput. Surv.},
	author = {Gardiner, Joseph and Nagaraja, Shishir},
	@date = {2016-12},
    year={2016},
	@note = {Place: New York, {NY}, {USA} Publisher: Association for Computing Machinery},
	keywords = {Command and control channels, botnets, data mining, machine learning, network intrusion},
}

@article{zhauniarovich_survey_2018,
	title = {A Survey on Malicious Domains Detection through {DNS} Data Analysis},
	volume = {51},
	@issn = {0360-0300},
	@url = {https://doi.org/10.1145/3191329},
	doi = {10.1145/3191329},
	abstract = {Malicious domains are one of the major resources required for adversaries to run attacks over the Internet. Due to the important role of the Domain Name System ({DNS}), extensive research has been conducted to identify malicious domains based on their unique behavior reflected in different phases of the life cycle of {DNS} queries and responses. Existing approaches differ significantly in terms of intuitions, data analysis methods as well as evaluation methodologies. This warrants a thorough systematization of the approaches and a careful review of the advantages and limitations of every group.In this article, we perform such an analysis. To achieve this goal, we present the necessary background knowledge on {DNS} and malicious activities leveraging {DNS}. We describe a general framework of malicious domain detection techniques using {DNS} data. Applying this framework, we categorize existing approaches using several orthogonal viewpoints, namely (1) sources of {DNS} data and their enrichment, (2) data analysis methods, and (3) evaluation strategies and metrics. In each aspect, we discuss the important challenges that the research community should address in order to fully realize the power of {DNS} data analysis to fight against attacks leveraging malicious domains.},
	number = {4},
	journaltitle = {{ACM} Comput. Surv.},
	author = {Zhauniarovich, Yury and Khalil, Issa and Yu, Ting and Dacier, Marc},
	date = {2018-07},
	@note = {Place: New York, {NY}, {USA} Publisher: Association for Computing Machinery},
	keywords = {Malicious domains detection, domain name system},
}

@article{metcalf_ecosystem_2021,
	title = {The Ecosystem of Detection and Blocklisting of Domain Generation},
	volume = {2},
	@url = {https://doi.org/10.1145/3423951},
	doi = {10.1145/3423951},
	abstract = {Malware authors use domain generation algorithms to establish more reliable communication methods that can avoid reactive defender blocklisting techniques. Network defense has sought to supplement blocklists with methods for detecting machine-generated domains. We present a repeatable evaluation and comparison of the available open source detection methods. We designed our evaluation with multiple interrelated aspects, to improve both interpretability and realism. In addition to evaluating detection methods, we assess the impact of the domain generation ecosystem on prior results about the nature of blocklists and how they are maintained. The results of the evaluation of open source detection methods finds all methods are inadequate for practical use. The results of the blocklist impact study finds that generated domains decrease the overlap among blocklists; however, while the effect is large in relative terms, the baseline is so small that the core conclusions of the prior work are sustained. Namely, that blocklist construction is very targeted, context-specific, and as a result blocklists do no overlap much. We recommend that Domain Generation Algorithm detection should also be similarly narrowly targeted to specific algorithms and specific malware families, rather than attempting to create general-purpose detection for machine-generated domains.},
	number = {3},
	journaltitle = {Digital Threats},
	author = {Metcalf, Leigh and Spring, Jonathan M.},
	@date = {2021-06},
    year={2021},
	@note = {Place: New York, {NY}, {USA} Publisher: Association for Computing Machinery},
	keywords = {{DGA}, {DNS}, blocklists, evaluation, metrics},
}

@article{georgoulias_botnet_2023,
	title = {Botnet Business Models, Takedown Attempts, and the Darkweb Market: A Survey},
	volume = {55},
	@issn = {0360-0300},
	@url = {https://doi.org/10.1145/3575808},
	doi = {10.1145/3575808},
	abstract = {Botnets account for a substantial portion of cybercrime. Botmasters utilize darkweb marketplaces to promote and provide their services, which can vary from renting or buying a botnet (or parts of it) to hiring services (e.g., distributed denial of service attacks). At the same time, botnet takedown attempts have proven to be challenging, demanding a combination of technical and legal methods, and often requiring the collaboration of a plethora of entities with varying jurisdictions. In this article, we map the elements associated with the business aspect of botnets and utilize them to develop adaptations of two widely used business models. Furthermore, we analyze the 28 most notable botnet takedown operations carried out from 2008 to 2021, in regard to the methods employed, and illustrate the correlation between these methods and the segments of our adapted business models. Our analysis suggests that the botnet takedown methods have been mainly focused on the technical side, but not on the botnet economic components. We aim to shed light on new takedown vectors and incentivize takedown actors to expand their efforts to methods oriented more toward the business side of botnets, which could contribute toward eliminating some of the challenges that surround takedown operations.},
	number = {11},
	journaltitle = {{ACM} Comput. Surv.},
	author = {Georgoulias, Dimitrios and Pedersen, Jens Myrup and Falch, Morten and Vasilomanolakis, Emmanouil},
	@date = {2023-02},
    year={2023},
	@note = {Place: New York, {NY}, {USA} Publisher: Association for Computing Machinery},
	keywords = {Cybercrime, attacks, botnets, business models, darkweb, economics, forum, marketplace, takedowns},
}

@online{krebs_china-based_2025,
	title = {China-based {SMS} Phishing Triad Pivots to Banks – Krebs on Security},
	url = {https://krebsonsecurity.com/2025/04/china-based-sms-phishing-triad-pivots-to-banks/},
	author = {Krebs, Brian},
	urldate = {2025-09-02},
	date = {2025-04-10},
	langid = {american},
}

@software{noauthor_salesforceja3_2025,
	title = {salesforce/ja3},
	rights = {{BSD}-3-Clause},
	url = {https://github.com/salesforce/ja3},
	publisher = {Salesforce},
	urldate = {2025-09-02},
	date = {2025-09-01},
	note = {original-date: 2017-06-13T22:54:10Z},
}

@online{noauthor_jarm_nodate,
	title = {{JARM}},
	url = {https://docs.censys.com/docs/ls-jarm},
	titleaddon = {Censys Documentation},
	urldate = {2025-09-02},
	langid = {english},
}

@online{noauthor_mar-10318845-1v1_2021,
	title = {{MAR}-10318845-1.v1 - {SUNBURST}},
	url = {https://www.cisa.gov/news-events/analysis-reports/ar21-039a},
	urldate = {2025-09-01},
	date = {2021-04-15},
	langid = {english},
}

@online{eckels_sunburst_nodate,
	title = {{SUNBURST} Additional Technical Details {\textbar} Mandiant},
	url = {https://cloud.google.com/blog/topics/threat-intelligence/sunburst-additional-technical-details},
	titleaddon = {Google Cloud Blog},
	author = {Eckels, Stephen and Smith, Jay and Ballenthin, William},
	urldate = {2025-09-01},
	langid = {english},
}

@software{wong_malwaremorghulissunburst_2024,
	title = {{MalwareMorghulis}/{SUNBURST}},
	rights = {{MIT}},
	url = {https://github.com/MalwareMorghulis/SUNBURST},
	author = {Wong, Adam Dorian},
	urldate = {2025-09-01},
	date = {2024-06-15},
	note = {original-date: 2022-03-01T15:06:29Z},
}

@article{thomson_novel_2025,
	title = {A Novel {TLS}-Based Fingerprinting Approach That Combines Feature Expansion and Similarity Mapping},
	volume = {17},
	@url = {https://www.ezproxy.dsu.edu/login?url=https://www.proquest.com/scholarly-journals/novel-tls-based-fingerprinting-approach-that/docview/3181453768/se-2?accountid=27073},
	doi = {10.3390/fi17030120},
	abstract = {Malicious domains are part of the landscape of the internet but are becoming more prevalent and more dangerous both to companies and to individuals. They can be hosted on various technologies and serve an array of content, including malware, command and control and complex phishing sites that are designed to deceive and expose. Tracking, blocking and detecting such domains is complex, and very often it involves complex allowlist or denylist management or {SIEM} integration with open-source {TLS} fingerprinting techniques. Many fingerprinting techniques, such as {JARM} and {JA}3, are used by threat hunters to determine domain classification, but with the increase in {TLS} similarity, particularly in {CDNs}, they are becoming less useful. The aim of this paper was to adapt and evolve open-source {TLS} fingerprinting techniques with increased features to enhance granularity and to produce a similarity-mapping system that would enable the tracking and detection of previously unknown malicious domains. This was achieved by enriching {TLS} fingerprints with {HTTP} header data and producing a fine-grain similarity visualisation that represented high-dimensional data using {MinHash} and Locality-Sensitive Hashing. Influence was taken from the chemistry domain, where the problem of high-dimensional similarity in chemical fingerprints is often encountered. An enriched fingerprint was produced, which was then visualised across three separate datasets. The results were analysed and evaluated, with 67 previously unknown malicious domains being detected based on their similarity to known malicious domains and nothing else. The similarity-mapping technique produced demonstrates definite promise in the arena of early detection of malware and phishing domains.},
	pages = {120},
	number = {3},
	journaltitle = {Future Internet},
	author = {Thomson, Amanda and Maglaras, Leandros and Moradpoor, Naghmeh},
	date = {2025},
	@note = {Place: Basel Publisher: {MDPI} {AG}},
	keywords = {Chemical fingerprinting, Command and control, Computers--Internet, Cybercrime, Literature reviews, Malware, Mapping, Metadata, Phishing, Privacy, Tracking, active fingerprinting, detection methods, malware domains, passive fingerprinting, phishing domains},
}

@article{guptta_modeling_2024,
	title = {Modeling Hybrid Feature-Based Phishing Websites Detection Using Machine Learning Techniques},
	volume = {11},
	@issn = {21985804},
	@url = {https://www.ezproxy.dsu.edu/login?url=https://www.proquest.com/scholarly-journals/modeling-hybrid-feature-based-phishing-websites/docview/2921812398/se-2?accountid=27073},
	doi = {10.1007/s40745-022-00379-8},
	abstract = {In this paper, we mainly present a machine learning based approach to detect real-time phishing websites by taking into account {URL} and hyperlink based hybrid features to achieve high accuracy without relying on any third-party systems. In phishing, the attackers typically try to deceive internet users by masking a webpage as an official genuine webpage to steal sensitive information such as usernames, passwords, social security numbers, credit card information, etc. Anti-phishing solutions like blacklist or whitelist, heuristic, and visual similarity based methods cannot detect zero-hour phishing attacks or brand-new websites. Moreover, earlier approaches are complex and unsuitable for real-time environments due to the dependency on third-party sources, such as a search engine. Hence, detecting recently developed phishing websites in a real-time environment is a great challenge in the domain of cybersecurity. To overcome these problems, this paper proposes a hybrid feature based anti-phishing strategy that extracts features from {URL} and hyperlink information of client-side only. We also develop a new dataset for the purpose of conducting experiments using popular machine learning classification techniques. Our experimental result shows that the proposed phishing detection approach is more effective having higher detection accuracy of 99.17\% with the {XG} Boost technique than traditional approaches.},
	pages = {217--242},
	number = {1},
	journaltitle = {Annals of Data Science},
	author = {Guptta, Sumitra Das and Shahriar, Khandaker Tayef and Alqahtani, Hamed and Alsalman, Dheyaaldin and Sarker, Iqbal H.},
	@date = {2024-02},
    year={2024},
	@note = {Place: Heidelberg Publisher: Springer Nature B.V.},
	keywords = {Anti-phishing, Business And Economics--Computer Applications, Cybercrime, Cybersecurity, Heuristic, Hybrid feature, Hyperlink feature, Machine learning, Phishing, Phishing detection, Real time, Search engines, Social security, {URL} feature, Websites, {XG} Boost},
}

@misc{wong_detecting_2023,
      title={Detecting Domain-Generation Algorithm {(DGA)} Based Fully-Qualified Domain Names {(FQDNs)} with Shannon Entropy}, 
      author={Adam Dorian Wong},
      year={2023},
      eprint={2304.07943},
      archivePrefix={arXiv},
      @primaryClass={cs.CR},
      @url={https://arxiv.org/abs/2304.07943}, 
}

@article{hassaoui_unsupervised_2024,
	title = {Unsupervised Clustering for a Comparative Methodology of Machine Learning Models to Detect Domain-Generated Algorithms Based on an Alphanumeric Features Analysis},
	volume = {32},
	@issn = {10647570},
	@url = {https://www.ezproxy.dsu.edu/login?url=https://www.proquest.com/scholarly-journals/unsupervised-clustering-comparative-methodology/docview/2918767893/se-2?accountid=27073},
	doi = {10.1007/s10922-023-09793-6},
	abstract = {Domain Generation Algorithms ({DGAs}) are often used for generating huge amounts of domain names to maintain command and control between the infected computer and the bot master. By establishing as needed a great number of domain names, attackers may mask their C2 servers and escape detection. Many malware families have switched to a stealthier contact approach. Therefore, the traditional methods become ineffective. Over the past decades, many researches have started to use artificial intelligence to create systems able to detect {DGA} in traffic, but these works do not use the same data to evaluate their models. This article proposes a comparative methodology to compare machine learning models based on unsupervised clustering and then applied this methodology to study the best models belonging to neural network methods and traditional machine learning methods to detect {DGAs}. We extracted 21 linguistic features based on the analysis of alphanumeric and n-gram, we studied the correlation between these features in order to reduce their number. We examine in detail those Machine learning algorithms and we discuss the drawbacks and strengths of each method with specific classes of {DGA} to propose a new switch case model that could be always reliable to detect {DGAs}.},
	pages = {18},
	number = {1},
	journaltitle = {Journal of Network and Systems Management},
	author = {Hassaoui, Mohamed and Hanini, Mohamed and El Kafhali, Said},
	@date = {2024-01},
    year={2024},
	@note = {Place: New York Publisher: Springer Nature B.V.},
	keywords = {Algorithms, Artificial intelligence, Clustering, Command and control, Computers--Computer Networks, Domain generation algorithm, Domain name, Domain names, Machine learning, Methodology, Neural networks},
}

@misc{wong2025gravity,
title={Gravity Falls},
author={Wong, Adam Dorian},
year={2025},
doi={10.5281/zenodo.17624554},
}

\end{document}